\documentstyle[prl,aps,epsf,rotate,multicol]{revtex}

\begin{document}

\title{Betweenness Centrality in Large Complex Networks}

\author{Marc Barth\'elemy}

\address{
CEA, D\'epartement de Physique Th\'eorique et Appliqu\'ee\\
BP12 Bruy\`eres-Le-Ch\^atel, France}

\maketitle
\begin{abstract}

We analyze the betweenness centrality (BC) of nodes in large complex
networks. In general, the BC is increasing with connectivity as a
power law with an exponent $\eta$. We find that for trees or networks
with a small loop density $\eta=2$ while a larger density of loops
leads to $\eta<2$. For scale-free networks characterized by an
exponent $\gamma$ which describes the connectivity distribution decay,
the BC is also distributed according to a power law with a non
universal exponent $\delta$. We show that this exponent $\delta$ must
satisfy the exact bound $\delta\geq (\gamma+1)/2$. If the scale free
network is a tree, then we have the equality $\delta=(\gamma+1)/2$.

\end{abstract}



\begin{multicols}{2}


\section{Introduction}

\begin{figure}
\narrowtext
\centerline{
\epsfysize=0.45\columnwidth{\epsfbox{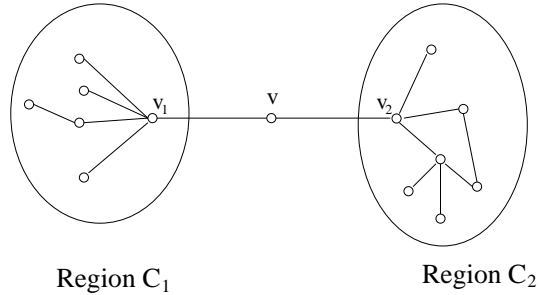}}
}
\vspace*{0.1cm}
\caption{ The node $v$ has a small connectivity (only two neighbors)
but all shortest paths from region $1$ to region $2$ has to go through
$v$ which implies a very large centrality. In fact, $v$ is here a
cut-vertex; its removal will break the network into two disconnected
components.}
\label{Simple_example}
\end{figure}

In large complex networks, not all nodes are equivalent.  For example,
the removal of a node can have a very different effect depending on
the node. If the node is at a dead-end, its removal will be without
any effect in contrast with the case of a cut-vertex (the analog of a
bridge for edges) which removal creates new disconnected
components\cite{Berge:1976,Clark:1991}. This question of the
importance of nodes in a network is thus of primary interest since it
concerns crucial subjects such as networks resilience to
attacks\cite{Albert:1999,Cohen:2001,Holme:2002} and also immunization
against epidemics\cite{Pastor-Satorras:2002}. In social network analysis, this
problem of determining the rank---or the ``centrality''---of the
actors according to their position in the social structure was studied
a long time ago\cite{Freeman:1977,Wasserman:1994}. Different
quantities were then defined in this context of social networks in
order to quantify this centrality. The simplest proxy for centrality
one could think of is the connectivity. However, the inspection of a
simple example such as the one in Fig.~\ref{Simple_example} shows that
centrality is in general not related to connectivity. The reason is
that connectivity is a local quantity which does not inform about the
importance of the node in the network. Indeed, the node $v$ in Fig.~1
has a small connectivity and the effect of its removal is not
determined by its connectivity but by the fact that it links together
different parts of the network. A good measure of the centrality of a
node has thus to incorporate a more global information such as its
role played in the existence of paths between any two given nodes in
the network. One is thus naturally led to the definition of the
betweenness centrality (BC) which counts the fraction of shortest
paths going through a given node. More precisely, the BC of a node $v$
is given by\cite{Freeman:1977,Wasserman:1994}
\begin{equation}
g(v)=\sum_{s\neq v\neq t}\frac{\sigma_{st}(v)}{\sigma_{st}}
\label{def_BC}
\end{equation}
where $\sigma_{st}$ is the total number of shortest paths from node
$s$ to node $t$ and $\sigma_{st}(v)$ is the number of shortest paths
from $s$ to $t$ going through $v$. In the following we will also use
the pair-dependency defined as\cite{Brandes:2001}
\begin{equation}
\mu_{st}(v)=\frac{\sigma_{st}(v)}{\sigma_{st}}
\end{equation}
The betweenness centrality $g$ scales as the number of pairs of nodes
($s\ne t\ne v$) and some authors rescale it by $(N-1)(N-2)/2$ in order
to get a number in the interval $[0,1]$ ($N$ is the number of nodes in
the giant component of the network). A naive algorithm for computing
$g$ would lead to a complexity of order ${\cal O}(N^3)$ and would thus
be prohibitive for large networks. Fortunately a rapid algorithm was
recently proposed\cite{Brandes:2001} which reduces the complexity to
${\cal O}(N^2)$ allowing the computation of the centrality for large
networks.

The definition (\ref{def_BC}) is indeed a good description of
centrality as can be easily seen on the example of
figure~\ref{Simple_example}. The BC of the node $v$ is given by
\begin{eqnarray}
g(v)&=&2\sum_{s\in C_1, t\in C_2}\frac{\sigma_{st}(v)}{\sigma_{st}}\\
&=&2\sum_{s\in C_1, t\in C_2}1\\
&=&2N_1N_2
\end{eqnarray}
where $N_1$ ($N_2$) is the number of nodes in region $C_1$ ($C_2$).
The first equality comes from the fact that the term for which $s$ and
$t$ are in the same region does not contribute since in this case
$\sigma_{st}(v)=0$. This result shows that although $v$ has a small
connectivity, its BC defined by (\ref{def_BC}) is large as intuitively
expected. This little argument prefigures the more general one about
centrality for trees (see below).

High values of the centrality thus indicate that a node can reach the
others on short paths or that this vertex lies on many short paths. If
one removes a node with large centrality it will lengthen the paths
between many pairs of nodes. The extreme case is when the node is a
cut-vertex\cite{Berge:1976,Clark:1991} and its removal creates new
connected components.  This was for example used in\cite{Huberman} to
determine recursively different communities in large networks.

There are other centrality indices based on shortest paths linking
pairs of nodes (stress, closeness, or graph
centrality\cite{Wasserman:1994,Brandes:2001}). In order to take into
account the fact that shortest paths are not always relevant, other
definitions were introduced such as the flow
betweenness\cite{Freeman:1991} and recently a betweenness centrality
based on random walks\cite{Newman:2003}. This definition
(\ref{def_BC}) differs from the following one which includes the paths
endpoints $s$ and $t$
\begin{equation}
\tilde{g}(v)=\sum_{s\neq t}\frac{\sigma_{st}(v)}{\sigma_{st}}
\end{equation}
($s$ or $t$ can be $v$). It can be easily checked that
\begin{eqnarray}
\tilde{g}(v)&=&\sum_{s=v\neq t}\mu_{st}(v)+\sum_{s\neq t=v}\mu_{st}(v)
+\sum_{s\neq v\neq t}\mu_{st}(v)\\
&=&2(N-1)+g(v)
\end{eqnarray}
This additional term $2(N-1)$ is sub-dominant since $g\sim {\cal
O}(N^2)$ and is thus negligible in the limit of large networks leading
to the same results for both definitions (for a typical value of the
order $N=10^4$, the relative difference for large connectivities is
negligible---of order $10^{-4}$---but could be larger for lower
$k$). In this work, we use the definition (\ref{def_BC}) and restrict
ourselves to non-weighted and non-directed graphs. We will rescale the
BC by $(N-1)(N-2)/2$ so that $g\in [0,1]$. We will keep the same
notation $g$ for this normalized centrality.

\section{Centrality and Connectivity}

It has been observed\cite{Amaral:2000} that large networks can be
essentially classified in two categories according to the decay of the
connectivity distribution $P(k)$.  The first category comprises the
``exponential'' networks with a connectivity distribution decaying
faster than any power law (random graph, Poisson graph, etc). In
contrast, the second category is constituted by the ``scale-free''
networks which have a probability distribution decaying as a power law
characterized by an exponent $\gamma$
\begin{equation}
P(k)\sim k^{-\gamma}
\end{equation}
For these networks, there are no typical nodes since the connectivity
can vary over a large range of values. In this sense, scale-free
networks are very heterogeneous compared to exponential networks for
which connectivity fluctuations are small.

In the following, we will investigate the BC for networks which are
simple models representative of each class.

\subsection{Scale-Free Networks}


In the case of scale-free networks, Goh et al have presented a
numerical study of the BC (or ``load'') distribution in a static
scale-free network model \cite{Goh:2001}. For this scale-free model,
the exponent $\gamma\in ]2,\infty[$ is a tunable parameter. They also
studied the scale-free model obtained by preferential
attachment\cite{Barabasi:1999} for which $\gamma=3$. They showed that
the BC is distributed according to a power-law with
exponent $\delta$ \cite{Note0}
\begin{equation}
P(g)\sim g^{-\delta}
\label{Pg}
\end{equation}
This behavior holds for large $g$ up to a cut-off value which is
controlled by finite-size effects. On the basis of their numerical
results, they conjectured that the value of $\delta\simeq 2.2$ is
``universal'' for all values of $\gamma\in ]2,3]$. Universality is
usually invoked in physics when different systems show the same
behavior\cite{Kadanoff}. For example many of the observed second order
phase transitions have a behavior which depends only on the dimension
of the system and the symmetry of the order parameter. In terms of the
renormalization group, all these systems are described by the same
fixed point of the renormalization group transformation and their
critical exponents are then equal. In the case of networks, Goh et
al\cite{Goh:2001,GohPNAS:2002} measured the exponent $\delta$ for
different real-world and in silico systems and found only two
classes\cite{GohPNAS:2002}: Either $\delta\simeq 2.2$ (Class I) or
$\delta=2$ (Class II). According to these numerical findings, they
claimed that there is ``universality'' and that networks could be
classified according to the value of $\delta$. This means that within
a given class, $\delta$ is independent of the details of the network
such as the mean connectivity $<k>=2m$, or the exponent $\gamma$.

\begin{figure}
\narrowtext
\centerline{
\epsfysize=0.45\columnwidth{\epsfbox{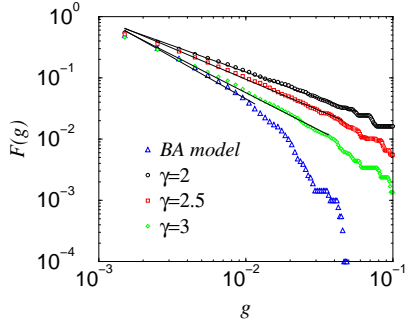}}}
\vspace*{0.5cm}
\caption{ Cumulative function of the load for different values of
$\gamma=2$, $2.5$, and $3$ (for $m=2$). These results were obtained
with the same values as in{\protect\cite{Goh:2001}} $N=10^4$ and for
$10$ configurations. The power law fits (straight lines) give the
values $\delta=1.86$, $2.01$ and $2.23$ while for the BA model
$\delta\simeq 2.3$.}
\label{Fg.and.bc}
\end{figure}

The value of $\delta$ is however not universal\cite{Barthelemy:2003}
and varies significantly as $\gamma$ changes in the interval $]2,3]$
or as $m$ varies. In order to see this non-universality, we first
computed the cumulative function $F(g)=\text {Prob}(\text{BC}\ge g)$
for the model proposed in\cite{Goh:2001} and for the scale-free
network obtained by preferential attachment\cite{Barabasi:1999}. The
results are shown on Fig.~\ref{Fg.and.bc} and even if the variations
are small, the differences are significant enough to show that
$\delta$ varies. However, as it can be seen on this
Fig.~\ref{Fg.and.bc} for the BA case, the power law is screened by a
cut-off which can be small due to finite-size effects.

\begin{figure}
\narrowtext
\centerline{
\epsfysize=0.45\columnwidth{\epsfbox{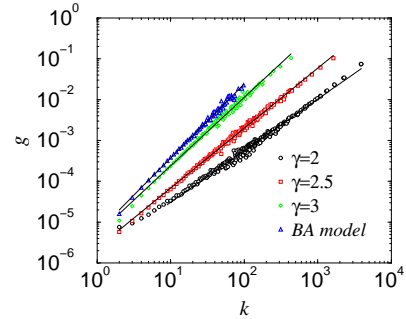}}
}
\vspace*{0.5cm}
\caption{ Log-Log plot of the normalized average load versus
connectivity for the same models as in {\protect\cite{Goh:2001}} with
$m=2$. The power law fits (straight lines) give $\eta=1.27\pm0.01$
($N=3.10^4$), $1.467\pm0.006$ ($N=5.10^4$), and $1.68\pm0.02$
($N=5.10^4$) for $\gamma=2$, $2.5$, and $3$ respectively. For the BA
model, $\eta=1.81\pm0.02$ ($N=5.10^4$).
}
\label{bc}
\end{figure}

The variations of $\delta$ obtained with $F(g)$ are significant enough
to claim that it is not a universal exponent but in order to
double-check our results we can also use an indirect way of computing
$\delta$. We study the relation between the load and the
connectivity\cite{Goh:2001,Vasquez:2002} which is of the form
\begin{equation}
g\sim k^\eta
\label{g.vs.k}
\end{equation}
where the exponent $\eta$ depends on the network. This relation
(between two random variables) implies that for a given value of $k$,
the corresponding value $g_k$ of the centrality is fixed. Due to noise
such as finite-size effects, $g_k$ can however have small fluctuations
and we compute the average of $g_k$ at fixed $k$. The result is shown
on Fig.~\ref{bc} and as can be seen on this plot, the power law
(\ref{g.vs.k}) holds remarkably for a large range of $k$ and allows an
accurate measure of $\eta$. In addition, this relation (\ref{g.vs.k})
enables us to estimate the cut-off value above which the power-law
(\ref{Pg}) does not hold. Indeed, the maximum connectivity scales
as\cite{Doro} $k_c\sim N^{1/(\gamma-1)}$ which thus implies that the
maximum BC scales as $g_c\sim N^{\eta/(\gamma-1)}$. Finally, we also
checked that the value of $\eta$ does not change significantly for
different values of the system size: For $\gamma=2.5$, we obtain
$\eta(N=10^4)=1.461\pm 0.005$, $\eta(N=2.10^4)=1.467\pm 0.006$, and
$\eta(N=5.10^4)=1.467\pm 0.006$ which represents a relative variation
due to size less than $1\%$).

The exponents $\eta$ and $\delta$ are not independent since
Eq.~(\ref{g.vs.k}) implies that
\begin{equation}
P(g)=\int dkP(k)\delta(g-k^{\eta})
\end{equation}
which for large $g$ implies a large $k$ and 
\begin{eqnarray}
\nonumber
P(g\gg 1)&\sim&\int dk k^{-\gamma}\delta(g-k^{\eta})\\
&\sim&g^{-1-\frac{\gamma-1}{\eta}}
\end{eqnarray}
which proves the following equality\cite{Vasquez:2002}
\begin{equation}
\eta=\frac{\gamma-1}{\delta-1}
\label{rel_expo}
\end{equation}
If the value of $\delta\simeq 2.2$ is
universal then $\eta$ is a linear function of $\gamma$ with slope
$\simeq 1/1.2\simeq 0.83$. 

\begin{figure}
\narrowtext
\centerline{
\epsfysize=0.45\columnwidth{\epsfbox{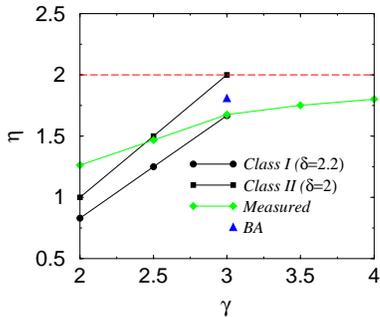}}
}
\vspace*{0.5cm}
\caption{ Exponent {\protect{$\eta$}} versus {\protect{$\gamma$}}. If
the universality proposed in {\protect\cite{Goh:2001}} would be
correct, the measured values for $\gamma\in[2,3)$ should lie on the
``universal'' straight line corresponding to $\delta=2.2$ (class I).}
\label{univ}
\end{figure}

In Fig.~\ref{univ} we plot the measured $\eta$ versus $\gamma$ for the
different types of networks studied and the corresponding value
predicted by universality. This Fig.~\ref{univ} shows that if for
$\gamma\simeq 3$ the value $\delta=2.2$ seems to be acceptable, the
claim of universality for $\gamma\in ]2,3]$ proposed in\cite{Goh:2001}
does not hold (our results do not fit in the other class $\delta=2.0$
either). In addition, we tested the universality for different values
of $m$ and we also obtain variations ruling it out: For $\gamma=2.5$
and for $N=2.\/10^4$, we obtain $\eta=1.477\pm0.006, 1.56\pm0.006$,
and $1.64\pm0.01$ for $m=2,4,6$ respectively. Even if Goh et al have
recently shown\cite{GohCondmat:2003} with a variant of the BA model
that for $m\in[1,2]$, the exponent $\delta$ is close to $2.2$ for
other models supposed to be within the same universality class (BA model,
static model, etc.), the exponent $\delta$ varies with $m$ or $\gamma$
and is therefore not universal.

We also note in Figure~\ref{univ} that for larger values of $\gamma$,
the exponent $\eta$ seems to converge to the value $\eta=2$. This
seems to show that for an exponential network, formally characterized
by $\gamma=\infty$, the exponent $\eta$ is equal to two. We will
discuss this fact in more details below.

Finally, the case $m=1$ for the preferential attachment is special in
the sense that the obtained scale-free network is a tree. Exact
calculations in this case\cite{Szabo:2002,GohPNAS:2002} show that
$\delta=2=\eta$. We will see below that the value $\eta=2$ is in fact
expected for any tree and that $\delta=2$ is the expected value for a
scale-free tree only with $\gamma=3$.

\subsection{Random graph}

\begin{figure}
\narrowtext
\centerline{
\epsfysize=0.45\columnwidth{\epsfbox{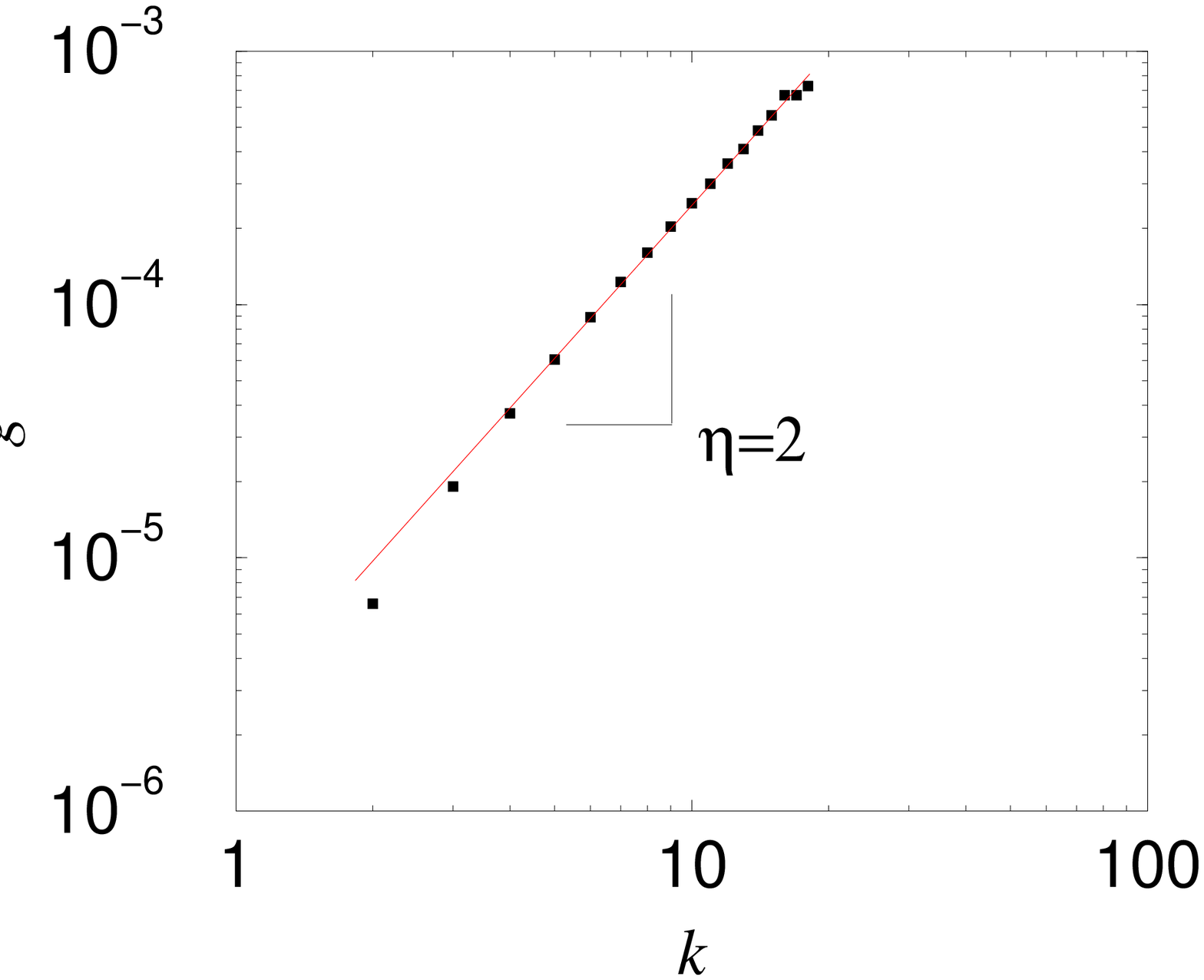}}
}
\vspace*{0.5cm}
\caption{ Log-Log plot of the normalized average load versus
connectivity for the random graph model with $N=5.10^4$ and
$<k>=6$). The straight line is of slope $\eta=2$.}
\label{bc_ER}
\end{figure}

We have seen different examples of scale-free networks in the previous
section and we focus now on the random
graph\cite{Bollobas:1985,Renyi:1980} (often called Erdos-Renyi graph)
which is a typical example of exponential networks for which the
connectivity distribution is decaying at least as fast as an
exponential. This network is constructed as follows. Starting from $N$
nodes, one connects with probability $p$ each pair of nodes. The
average final number of edges is thus $E=pN(N-1)/2$ and the average
connectivity is $2E/N=p(N-1)\simeq pN$ for large graphs. More
generally, the probability that a node has connectivity $k$ is given
by the Binomial law
\begin{equation}
p(k)=\left(^{N-1}_{\;\;\;k}\right) p^k(1-p)^{N-1-k}
\end{equation}
which converges to a Poisson law of parameter $<k>$ for large $N$ and
small $p$ such that $<k>=pN$ is fixed.  We studied the centrality for
this network and in Fig.~\ref{bc_ER} we plot the measured BC versus
the connectivity. Even if the connectivity is not varying over a very
large range, this plot shows that for large $k$ we have $\eta=2$. We
will discuss this result in more details below but we already note
that the random graph has a very small clustering coefficient $C\sim
1/N$ ($C$ counts the average fraction of pairs of connected
neighbors\cite{Watts:1998}) and that this property could possibly be
related to the fact that $\eta=2$.

\section{Discussion and Analysis of the results}

The results obtained above show that the exponents
$\eta$ and $\delta$ are not universal and depend on the details of the
network. In particular, if the network is scale-free (tree-like or not)
$\delta$ depends on the exponent $\gamma$ which describes the power law
decay of the connectivity distribution.

The important exponent appears to be $\eta$ which describes how the
betweenness centrality depends on the connectivity. The ``optimal''
situation which maximizes the BC for a vertex is obtained when all shortest
paths are going through it, which happens for a tree structure (ie. a
network without loops). To this optimal tree situation corresponds the
maximum value of $\eta=2$. In order to show this, we first define some
objects. If a vertex $v$ has connectivity $k$, we denote by $v_i$
($i=1,\dots,k$) its $k$ neighbors. Each neighbor $v_i$ defines a
``neighborhood'' $C_i$ constituted by nodes which are closer to this
neighbor than to any other one. More formally, $C_i$ is defined as
follows
\begin{equation}
C_i=\{s\; |\;d(s,v_i)\le d(s,v_j)\;\forall j\neq i\}
\end{equation}
When the equality of distances $d(s,v_i)=d(s,v_j)$ is obtained for
some $j$ then the node $s$ belongs to the two neighborhoods $C_i$ and
$C_j$. The existence of a non empty intersection between different
neighborhoods allows for the possibility of paths by-passing
the node $v$.

\begin{figure}
\narrowtext
\centerline{
\epsfysize=0.5\columnwidth{\epsfbox{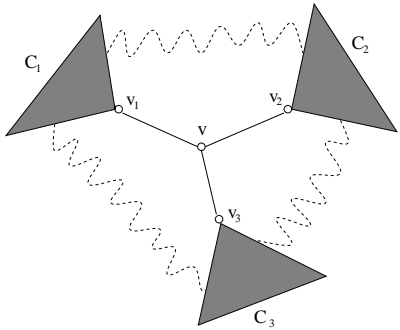}}
}
\vspace*{0.5cm}
\caption{ The node $v$ has here $3$ neighbors $v_1,v_2,v_3$. These
neighbors define three different regions which are disconnected in the
case of a tree. When the intersection of these regions is not empty,
(shortest) paths between these regions which by-pass $v$ can exist
(and are represented by the dotted line between the regions $C_i$).  }
\label{Comm}
\end{figure}

In the following we denote by $N_i$ the size of each region $C_i$. In
general the shortest paths from $s\in C_i$ to $t\in C_j$ go through
$v$ or avoid $v$ by using paths on nodes belonging to $C_i\cap C_j$
[see Fig.~\ref{Comm}]. If the two nodes $s$ and $t$ belong to the same
neighborhood, say $C_l$, there is always a shortest path within $C_l$ (in the
worst case the shortest path goes through $v_l$ but not through
$v$) and therefore
\begin{equation}
\mu_{st}(v)=0\;\;{\text if}\;\; s,t\in C_l
\end{equation}
In terms of these neighborhoods $C_i$, the BC can be rewritten as
\begin{eqnarray}
\nonumber\\
g(v)&=&\sum_{s\neq v\neq t}\mu_{st}(v)\\
&=&\sum_{i\neq j}\sum_{s\in C_i, t\in C_j}\mu_{st}(v)
\end{eqnarray}
(the term $i=j$ gives zero).

For a tree, these regions $C_i$ are disconnected one from the other
and the BC can then be rewritten as
\begin{equation}
g(v)\sim\sum_{i\neq j} N_iN_j
\end{equation}
If in addition these different parts are of the same order of
magnitude $N_i\simeq N_0$ (which is similar to a statistical isotropy
condition) we obtain
\begin{equation}
g(v)\sim N_0^2 k(k-1)
\end{equation}
which for large $k$ behaves as $k^2$ leading to the value
$\eta=2$. Obviously, the ``isotropy'' condition $N_i\simeq const.$ is
necessary and if it is not satisfied then the preceding argument does
not apply\cite{notes1}. We note that an exactly solvable model for
which this assumption is satisfied is the tree graph obtained with the
BA model with $m=1$ and where one indeed finds
$\eta=2$\cite{GohPNAS:2002}. The tree situation miximizes the BC since
all shortest paths are going through the node $v$. In any other cases,
the centrality will be less and the maximum possible value of $\eta$
is $2$. More generally, if for a network the density of loops is small
enough such that most shortest paths which go from $C_i$ to $C_j$ have
to go through $v$ then we obtain $\eta=2$. This is the case for trees
but also for random graphs for which the clustering is small $\sim
1/N$.

If in addition to be a tree, the network is scale-free we can use
the relation (\ref{rel_expo}) which together with $\eta=2$ leads to
\begin{equation}
\eta=2\Rightarrow\;\delta=\frac{\gamma+1}{2}
\label{treeSF}
\end{equation}
This relation in particular implies that for the scale-free BA network
with $m=1$ and $\gamma=3$, we obtain $\delta=(\gamma+1)/2=2$ in
agreement with previous results\cite{Szabo:2002,GohPNAS:2002}. It
should be noted that in both these
papers\cite{Szabo:2002,GohPNAS:2002} the authors demonstrate that
$\delta=2$ in the specific case of preferential attachment. However,
in\cite{GohPNAS:2002}, the authors claim that their result is valid
for any scale-free tree with $\gamma>2$. This is an incorrect
statement since their derivation is only valid for preferential
attachment and in general $\delta$ depends on $\gamma$ as predicted by
Eq.~(\ref{treeSF}).

On the other hand---and this is the second possible category of
networks---if there is a significant fraction of shortest paths which
by-pass $v$ then the exponent $\eta$ will be less than $2$. If the
network is scale-free then we can use the relation (\ref{rel_expo})
which together with $\eta<2$ leads to the exact bound
\begin{equation}
\eta<2\Rightarrow\;\delta>\frac{\gamma+1}{2}
\label{bound}
\end{equation}

The quantity $2-\eta$ is thus a measure of the density of loops in the
network. The fact that $\eta<2$ indicates that the different parts are
also connected by shortest paths which do not pass through the central
node. More generally, it would be interesting to understand how $\eta$
depends on the different parameters of the network such as $\gamma$,
the clustering coefficient, the loop density, the ``anisotropy'', or
any other correlation function.


In summary, it seems that concerning the betweenness centrality, we
can distinguish two main categories. For the first one which comprises
the trees and tree-like networks (clustering almost zero, density of
loops very small), we have $\eta=2$. If in addition, the tree is
scale-free with exponent $\gamma$, we have the relation
$\delta=(\gamma+1)/2$. The second category comprises the networks for
which the density of loops is large enough so that the networks are
very different from trees. In this case, the exponents
$\delta,\eta$---when they exist---are not universal and depend on the
different details (average connectivity, correlations, etc). If this
``clustered'' network is scale-free with exponent $\gamma$, the
exponent $\delta$ must obey an exact bound
[Eq.~(\ref{bound})]. Although we believe that the present picture is
the correct one, further studies are still necessary to understand
which are exactly the parameters which control the behavior of
$\eta$. In this respect, analytical insights would be particularly
valuable.

Acknowledgments: I thank the department of physics-INFN in Torino for
its warm hospitality during the time this work was started and the
Equipe R\'eseaux, Savoirs $\&$ Territoires at the Ecole Normale
Sup\'erieure, Paris.


\end{multicols}



\begin{references}
\bibitem{notes1} It would be interesting to quantify for different
types of networks the degree of anisotropy---measured by the
$N_i$'s---versus the connectivity.
\end{references}
\end{document}